\documentstyle[aps]{revtex}

\begin{document}

\title{Phase Diagram of a Superconducting and Antiferromagnetic 
System \\
with SO(5) Symmetry} 

\author{Xiao Hu,\(^1\) Tomio Koyama,\(^2\) and Masashi Tachiki\(^1\)}

\address{\(^1\)National Research Institute for Metals, 
Tsukuba 305-0047, Japan\\
\(^2\)Institute for Materials Research, Tohoku University,
Sendai 980-8577, Japan}

\date{submitted on Aug. 27, 1998}

\maketitle

\begin{abstract}
Temperature vs. chemical-potential phase diagrams of an SO(5) model 
for high-\(T_c\) cuprates are calculated by Monte Carlo simulation.  
There is a {\it bicritical point} where the second-order antiferromagnetism
 (AF) and superconductivity transition lines merge tangentially 
into a first-order line, and {\it the SO(5) symmetry is achieved}. In an 
external magnetic field, the AF ordering is 
first order in the region where the first-order melting line of 
flux lattice joins in.  There is a {\it tricritical point} on the AF 
transition line from which the AF ordering becomes second order.

\vskip1cm

\end{abstract}

\newpage

The antiferromagnetism (AF) and superconductivity (SC) exist
near to each other in the temperature vs. doping-rate phase diagrams of
many high-\(T_c\) cuprates.  It is understood that in 
these materials the same electrons can contribute either to 
AF or to SC depending on the doping rate of
holes, or the chemical potential of electrons.  It is then natural and 
important to  incorporate these two orders into a single theoretical
scheme.  Zhang proposed a unified theory based on the SO(5) 
symmetry of SC and AF  \cite{Zhang}, in which the three  
components of a spin, and the real and imaginary part of the SC
order parameter compose a five-component superspin; the chemical
potential plays the role of a symmetry-breaking field.  
Studies on the SO(5) model are carried out on the symmetry of quantum
operators by Rabello {\it et al}. \cite{Rabello} and on its relation with 
the long studied Hubbard and \(t-J\) models by  Meixner {\it et al}. and Eder {\it et al}. \cite{Meixner}. 
Using the SO(5) model Arovas {\it et al}. showed that the vortex line can 
have an AF core \cite{Arovas}.  A modified SO(5) vector pseudospin 
model is proposed and studied by Koyama \cite{Koyama}.
A systematic thermodynamic study on the SO(5) model is still lacking,
which is important because the entropy effect in this model is
significant, and ultimately one should compare the predictions by the model with phase diagrams observed experimentally.  To reveal the 
thermodynamic behavior of the SO(5) model is also of interests
from the viewpoint of the theory for phase transitions and critical
phenomena.  As shown in the following, the high-\(T_c\) cuprates may
be a new class of materials which show multicriticality, like magnets
and liquid crystals.

In this Letter, we report on our Monte Carlo (MC) study of a classic 
version of the SO(5) model.  Our main results are as follows: 
There are normal (N), AF, SC and phase-separation (PS) phases
in the temperature vs. chemical-potential phase diagram.  The 
N-AF and N-SC phase transitions are second order in the 3D 
Heisenberg and \(XY\) universality classes, respectively, while those between the ordered phases are first order.  The N-AF and N-SC lines merge tangentially at a {\it bicritical point} of a finite temperature
into the first-order AF-SC line on which the PS-AF and PS-SC lines 
collapse in the vicinity of the bicritical point.   
{\it By checking the weights of the SC and AF components
it is confirmed  in the present simulation that the SO(5) symmetry is
achieved at the bicritical point. }  In the presence of an 
external magnetic field,  the N-AF phase transition is
 first order in the vicinity of the point where the N-AF line meets the 
first-order melting line of the flux lattice (N-SC line).   {\it There is a 
tricritical point on the N-AF line from which the phase transition 
becomes second order.  
 This change in the order of phase transition is  
originated from the thermal fluctuations enhanced by the SC degrees of freedom, and is an important implication of the SO(5) theory.}

Generally speaking the thermodynamic nature of a phase transition is 
governed by the symmetry of the degrees of freedom, dimensionality, and
range of interactions.  Quantum effects are less important as far as the phase transition occurs at a finite temperature.  Therefore, in the 
present study we adopt the following classical SO(5) Hamiltonian on
 the simple cubic lattice:

\begin{equation}
 {\cal  H}
  =-J \sum_{\langle i,j\rangle}\cos\theta_i\cos\theta_j
           \cos(\varphi_i-\varphi_j)
    +J\sum_{\langle i,j\rangle}\sin\theta_i\sin\theta_j
          {\bf S}_i\cdot{\bf S}_j
     +g\sum_i\sin^2\theta_i
\end{equation}

\noindent with \(J>0\), \({\bf S}^2_i=1\).
The first term is for the SC order parameters where \(\varphi\)'s
are the phase variables, reminding that a superconducting 
transition is expected to belong to the 3D \(XY\) universality class.  The 
second term is for the AF components.    The parameter \(g\) is a 
renormalized chemical potential \cite{Zhang,Arovas}, and a
negative \(g\) favors AF components.  Hamiltonian (1) is equivalent
to the classical version of the SO(5) vector pseudspin model presented
in Ref.\cite{Koyama}. 
 At each MC step, a trial configuration of superspins is generated with  
\(\sin^2\theta=1/2\) in average and is then subjected to the standard 
Metropolis prescription under Hamiltonian (1).  This process reduces 
significantly the simulation time for generating symmetric 5-vector 
superspins where each component should have a weight of 1/5 in 
average.  The loci of the bicritical and tricritical points in phase diagrams may be shifted somehow to the negative \(g\) direction.
However, the important 
features, such as the properties around the bicritical point and the
existence of a tritritical point in an external magnetic field, should 
remain unchanged since they are determined by the symmetry of the
Hamiltonian.  Periodic boundary conditions are put in all the directions.  

The temperature vs. chemical-potential phase diagram is depicted in 
Fig. 1 using a system of size \(L^3=40^3\).  A bicritical point 
 \cite{Fisher} is observed at \([g_b, T_b]=[-1.04J, 0.845J/k_B]\), where
the N-AF and N-SC lines merge 
tangentially.  {\it By checking the weights of the SC and AF components
it is confirmed  that the SO(5) symmetry is recovered at the bicritical
 point. } 
The critical temperature for the SC ordering, \(T_{XY}(g) \), is higher 
than that of the AF ordering, \(T_N(g) \), for the same value of 
\(|g-g_b|\). These two phase transitions are in the 3D \(XY\) and 
Heisenberg universality classes, respectively, in spite of the existence
 of the competing degrees of freedom.  
Singularities in the temperature dependence of the weight 
\(\langle\sin^2\theta\rangle\) are observed only at the phase 
boundaries. Therefore,
no crossover temperatures such as \(T_s\) and \(T_p\) in 
Ref. \cite{Zhang} can be defined.  The superspin flips are incomplete even
at the phase transitions at finite temperatures in the sense
\(0<\langle \sin^2\theta\rangle <1\).

The segment between \([g_b, T_b]\) and \([0, 0]\) is considered as the 
first order SC-AF phase 
boundary.  This first-order phase transition is the counterpart of the
spin-flop transition in the uniaxially anisotropic antiferromagnets in a
 magnetic field parallel to the easy axis, as first pointed out by N\'eel 
\cite {Neel} and investigated by others \cite{Fisher,Aharony}. 
In the simulations, we observed the following hysteresis 
behaviors:  Heating the 
system from an AF configuration with \(\sin\theta=1\) 
at zero temperature, we observe a first-order AF to SC phase transition 
at \(T_{ht}(g)\)  when \(g_b<g<0\); Cooling the 
system in the SC phase, a first-order 
SC to AF transition occurs at \(T_{cl}(g)\) when \(g_b<g<-0.6 J\), with 
\(T_{cl}(g)<T_{ht}(g)\).   In the region \(g>-0.6 J\), the SC order survives
down to zero temperature.  
This hysteresis phenomenon suggests the coexistence of the SC and AF orders, presumably in the form of phase separation, in the parameter 
region shown in
Fig. 1. The PS phase shrinks as the chemical potential decreases and 
approaches \(g_b\).  Near the bicritical point \([g_b, T_b]\),  the \(T_{cl}
(g)\) and \(T_{ht}(g)\) lines collapse into a single line which is 
tangential to \(T_{XY}(g) \) and \(T_N(g) \) lines
at the bicritical point. Increasing \(g\) from \(g_b\), the latent heat 
associated with \(T_{ht}(g)\) increases from zero, assumes its maximum
 \(Q\simeq 0.2 J\) at \(g\simeq -0.8 J\), then  decreases and becomes 
zero at \(g=0\); the one associated with \(T_{cl}(g)\) increases linearly 
from zero to \(Q\simeq 0.6 J\) at \(g\simeq -0.6 J\)

In order to see the effect of anisotropy in the couplings, we have also simulated the Hamiltonian with \(\Gamma^2=J/J^{SC}_c=10\) ( for the
definition of \(\Gamma\) see Hamiltonian (2) in below).  
The phase diagram is shown in the right inset of Fig. 1 using a system of
size \(L^3=20^3\).  \(T_{XY}(g) \) becomes lower than \(T_N(g) \) for the
 same value of \(|g-g_b|\), as in
experimental observations of high-\(T_c\) cuprates.   The bicritical 
point is at \([g_b, T_b]=[0.23J, 0.65J/k_B]\) for this anisotropic case.  

An external magnetic field will couple with the AF moments 
in the Zeeman form, and meanwhile modify the phases of 
SC order parameters.  In an external magnetic field along the \(c\) axis
(\(\hat{c} \parallel \hat{z}\)), the Hamiltonian should be 
 
\begin{eqnarray}
{\cal H}=-J&&\left[\sum_{\langle i,j\rangle\parallel ab {\rm plane}}
        \cos\theta_i\cos\theta_j\cos\left(\varphi_i-\varphi_j
                -\frac{2\pi}{\phi_0}
                      \int^{j}_{i}{\bf A}^{(2)}\cdot d{\bf r}^{(2)}\right)
+\frac{1}{\Gamma^2}\sum_{\langle i,j\rangle\parallel c {\rm axis}} 
                     \cos\theta_i\cos\theta_j\cos(\varphi_i-\varphi_j)\right]
\nonumber \\
+J&&\sum_{\langle i,j \rangle}\sin\theta_i\sin\theta_j
      {\bf S}_i\cdot{\bf S}_j
    -H\sum_i\sin\theta_i S_{iz}+g\sum_i\sin^2\theta_i.
\end{eqnarray}

\noindent 
This Hamiltonian can be derived from the Ginzburg-Landau (GL)
Lawrence-Doniach (LD) free energy functional \cite{Teitel,Hu} with the 
additional AF components.  The GL free energy functional was used by 
Arovas {\it et al}. \cite{Arovas} to discuss the AF vortex core. 
In the present study we adopt \(f=1/25\) which means that there is
one flux line per 25 unit cells in the \(ab\) plane.  The system size is 
\(L_x\times L_y\times L_z=50\times 50\times 40\), and there are
100 flux lines induced by the external magnetic field in the system.
The anisotropy constant is taken as \(\Gamma^2=10\).  The magnetic 
induction is evaluated as \(B=10\phi_0/[25(\gamma d)^2]\) for a  
high-\(T_c \) cuprate of the distance between neighboring CuO\(_2\) 
bilayers \(d\) and the material-dependent anisotropy constant 
\(\gamma\)  \cite{Hu}.  We have put \(H=0.1\) in the Zeeman term, and
confirmed that this field confines the AF components
 almost perfectly in the \(ab\) plane.  The value of \(H\) is irrelevant to
the phase transitions and the phase diagram shown below, till it
increases to a critical value to cause the spin-flop transition in the AF
components.  This aspect will be discussed elsewhere.

The phase diagram for \(\Gamma^2=10\) is shown in Fig. 2.  The N-SC 
phase transition is the well established, first-order melting of flux-line lattice \cite{ZS}.  Cooling the system across the melting point 
\(T_m(g)\), 
the helicity modulus sets up sharply from zero; a \(\delta\)-function 
peak in the specific heat is detected associated with a tiny latent heat 
of order of  \(0.001 J\) per site; the flux lines manifest themselves
into the Abrikosov lattice; the hysteresis loop associated with this
first-order melting is very small. 
These observations are the same as those in 
absence of the AF degrees of freedom \cite{Hu}.  
The cores of flux lines are of finite AF components in the present case, 
as discussed by Arovas {\it et al}. \cite{Arovas}.  

The effect of the competition between the AF and SC degrees of freedom in the same Hamiltonian is most pronounced in the region around 
\([g_{\rm x}, T_{\rm x}]=[0.56 J, 0.41 J/k_B]\) where the N-AF line 
meets the first-order N-SC line. Namely, the N-AF phase transition is 
first order as seen in Fig. 3 for a cooling process of \(g=0.5J\).
Hysteresis behaviors are observed as denoted by \(T_{Nht}(g)\) and 
\(T_{Ncl}(g)\) in Fig. 2.
The change of the usual second-order AF ordering to first order is
because of the big thermal fluctuations enhanced by the SC degrees of 
freedom in the external magnetic field.  There is a tricritical point 
\cite{review,Aharony} on the N-AF line at 
\([g_t, T_t]=[0.40 J, 0.59 J/k_B]\). 
The temperature dependence of the staggered magnetization and specific
heat, as well as the weight of AF components are depicted in Fig. 4
for \(g=g_t\).  Presuming single power-law singularities and using the 
data in the critical region \(0.490\le k_B T/J\le 0.584\), we obtain 
the tricritical exponents \(\beta=0.24\pm 0.02\) for the staggered 
magnetization and \(\alpha^\prime=0.47\pm 0.06\) for the specific heat 
below the tricritical point, and the tricritical point 
\(k_BT_t/J=0.59\pm 0.01\).  For the specific heat above the tricritical 
point one clearly has \(\alpha=0\) from Fig. 4.  
These tricritical exponents are consistent with those in literature
\cite{review}.  Discussions about the logarithmic corrections
to the power-law functions suggested by the renormalization group
will be given elsewhere.  We notice that {\it the first-order AF
 ordering for \(g_t<g<g_{\rm x}\) and the existence of the 
tricritical point on the N-AF line are important implications of 
the SO(5) theory}. 
The AF ordering at \(T_N(g)\) switches back to a second-order transition 
for \(g<g_t\).  The phase transition from the normal phase to the 
Abrikosov 
SC phase is always first order, reflecting the fact that the SC order 
parameters in an external magnetic field have symmetry different from 
a simple 2-vector model.  Our simulated 
phase diagram in Fig. 2 is hence not included in the list of possible phase
diagrams by mean-field theory and renormalization group 
\cite{Aharony}. 

In summary we have computed the phase diagram of a classical version 
of Zhang's SO(5) model.  We have found a bicritical point
at a finite temperature where the second-order AF (3D Heisenberg class)
and SC (3D \(XY\) class) transition lines merge tangentially into a 
first-order line.  By checking the weights of the SC and AF components, we have observed the SO(5) symmetry at the 
bicritical point. Phase separation 
between the AF and SC orders is observed as the first-order line splits
into the supercooling and superheating lines.  In the external magnetic 
field along the \(c\) axis, we have found a tricritical point  
on the AF transition line where the phase transition changes from 
second order into first order for larger chemical potentials.
Quantum fluctuations in the AF components of spin \(S=1/2\) are not
included in the present calculation which may reduce the bicritical 
point.  A quantum Monte Carlo simulation on the quantum SO(5) model is 
hence an interesting future problem.  In order to
compare quantitatively the simulated phase diagrams with the
temperature vs. carrier-concentration phase diagrams for high-\(T_c\)
cuprates, or the temperature vs. pressure phase diagrams for organic
superconductors, one should reduce the SC coupling constants from the 
value of the AF ones.

Xiao Hu would like to thank S.-C. Zhang for stimulating and fruitful 
discussions. He also appreciates Dr. Y. Nonomura for a helpful discussion
 on determination of the tricritical exponents.
The present simulation is performed on the Numerical Materials 
Simulator (SX-4) of National Research Institute for Metals (NRIM), Japan.

%\newpage

\vskip3cm

\noindent Figure Captions

\noindent Fig. 1: Temperature vs. chemical-potential phase diagram of 
the SO(5) model. 
N: normal phase; AF: antiferromagnetic ordered phase;
SC: superconducting phase; PS: AF and SC phase-separation phase.
The mark \(b\) denotes the bicritical point.
The left inset shows the zoomed-in phase diagram around the 
bicritical point.
The right inset displays the phase diagram for the anisotropic case
of \(\Gamma^2=J/J^{SC}_c=10\).  

\noindent Fig. 2: Temperature vs. chemical-potential phase diagram of 
the SO(5) model in an external magnetic field.  
The mark \(t\) denotes the tricritical point.  SC denotes 
the superconducting Abrikosov flux-line-lattice phase.
The magnetic field in the Zeeman term is \(H=0.1J\).

\noindent Fig. 3: Temperature dependence of the internal energy,  
specific heat (A); staggered magnetization \({\bf m}_{\rm stag}
=\langle{\bf S}\sin\theta\rangle_{\rm stag}\), weight of AF 
components, 
and helicity modulus along the \(c\) axis (B), for a cooling process of 
\(g=0.5J\) where a first-order AF ordering is observed. The magnetic
 field in the Zeeman term is \(H=0.1J\).

\noindent Fig. 4: Temperature dependence of the staggered magnetization
\({\bf m}_{\rm stag}=\langle{\bf S}\sin\theta\rangle_{\rm stag}\),
weight of AF components and specific heat at \(g_t=0.40J\) where the 
tricriticality is observed. The magnetic field in the Zeeman term is 
\(H=0.1J\).

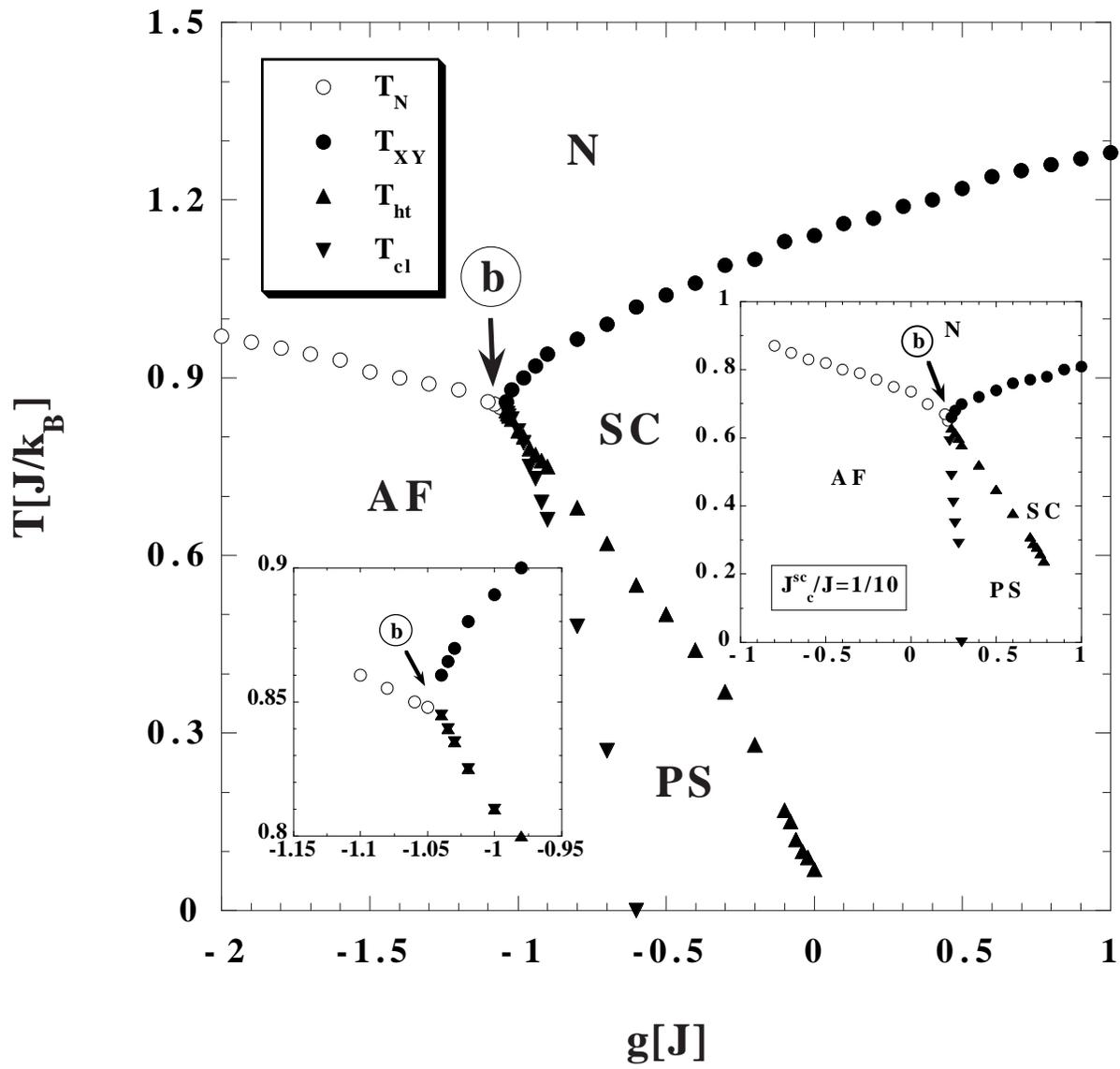
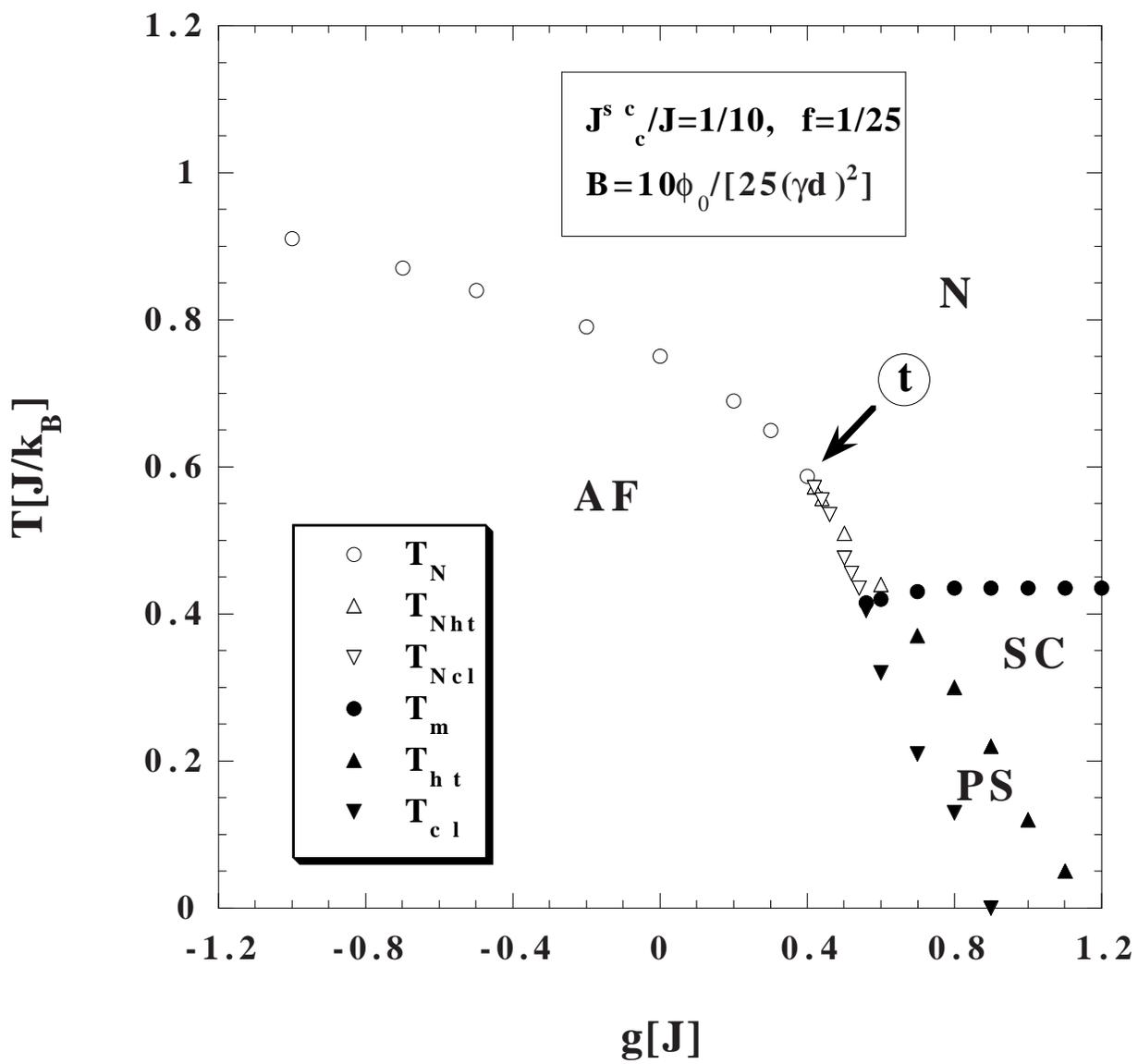
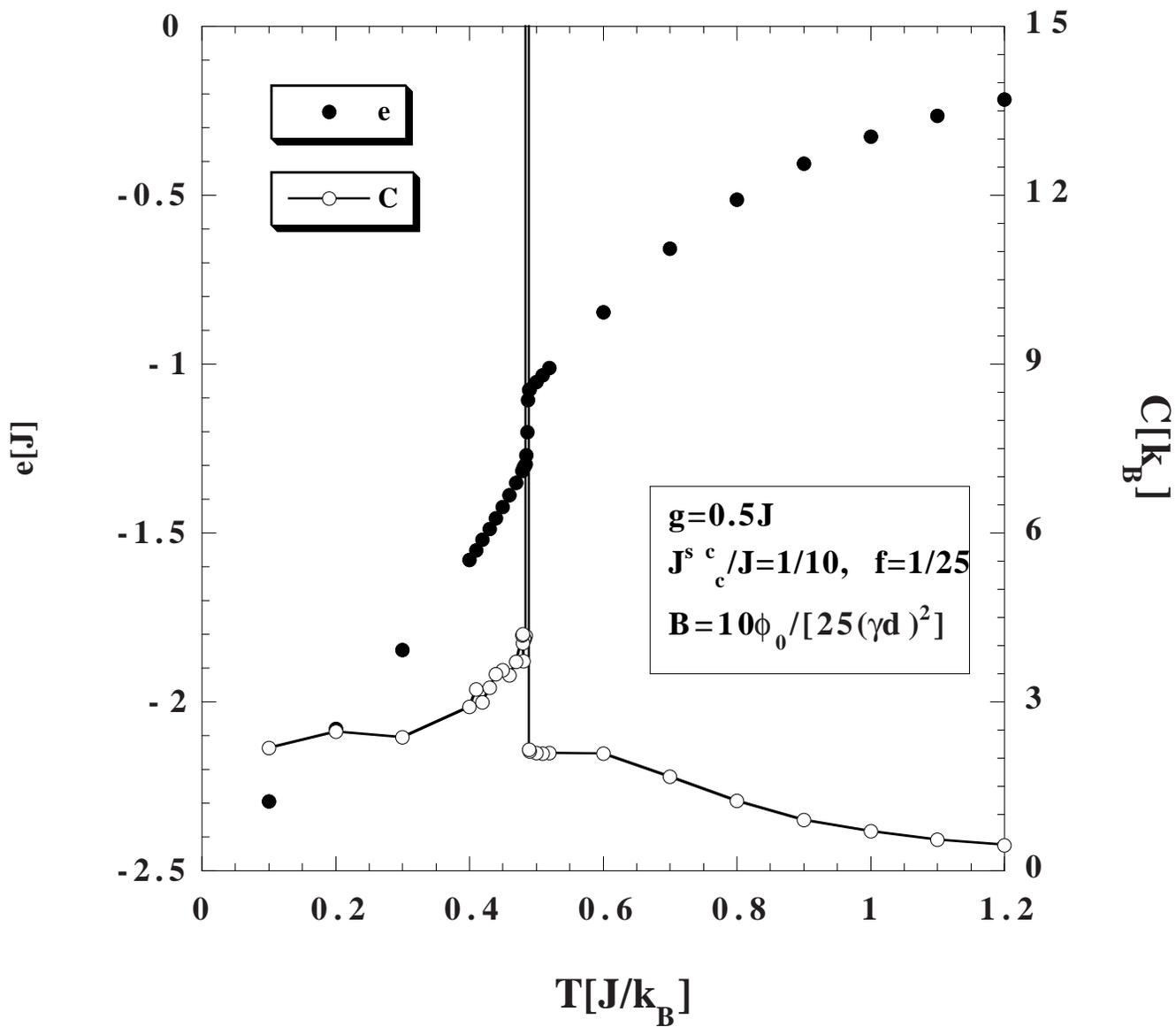
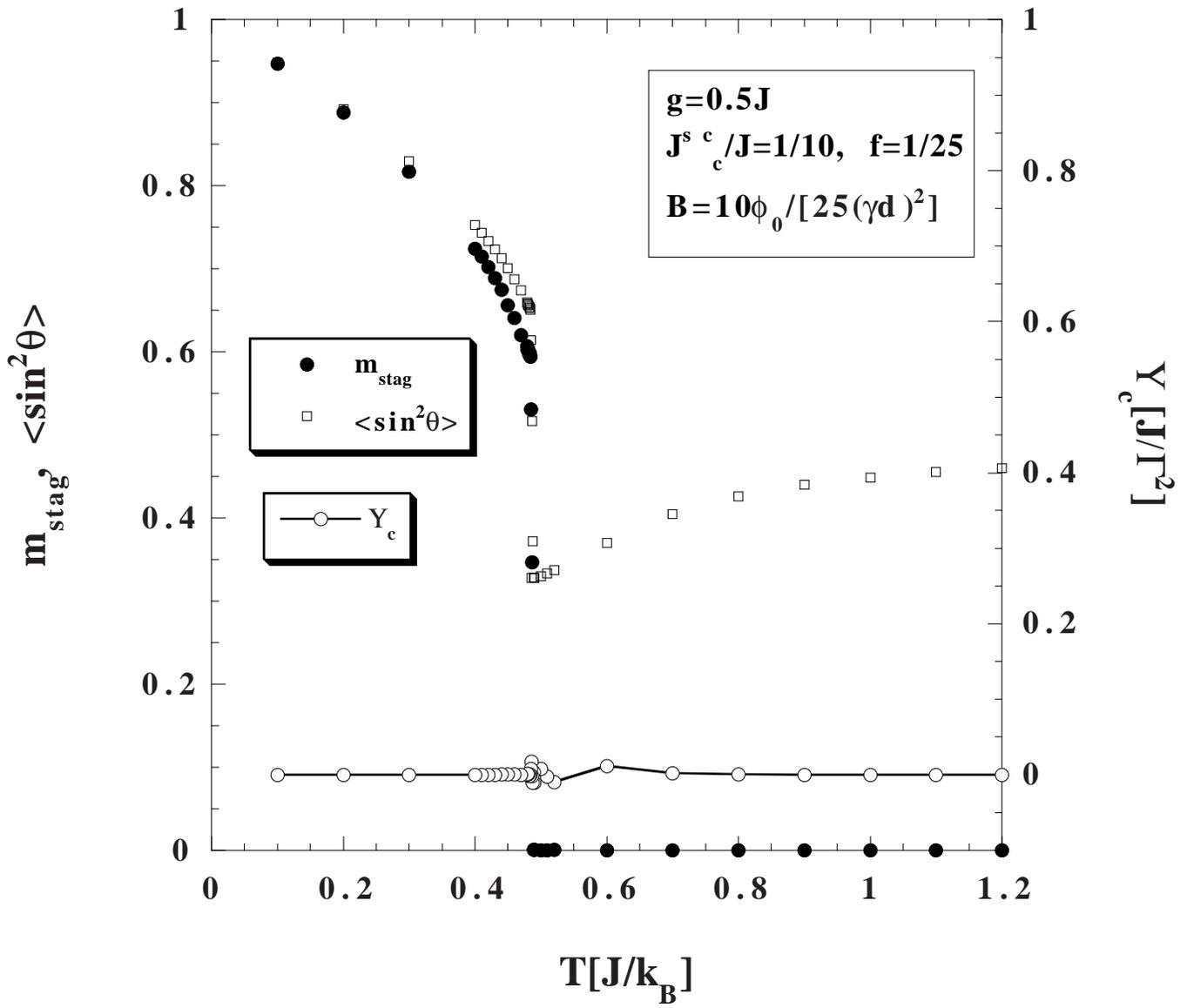
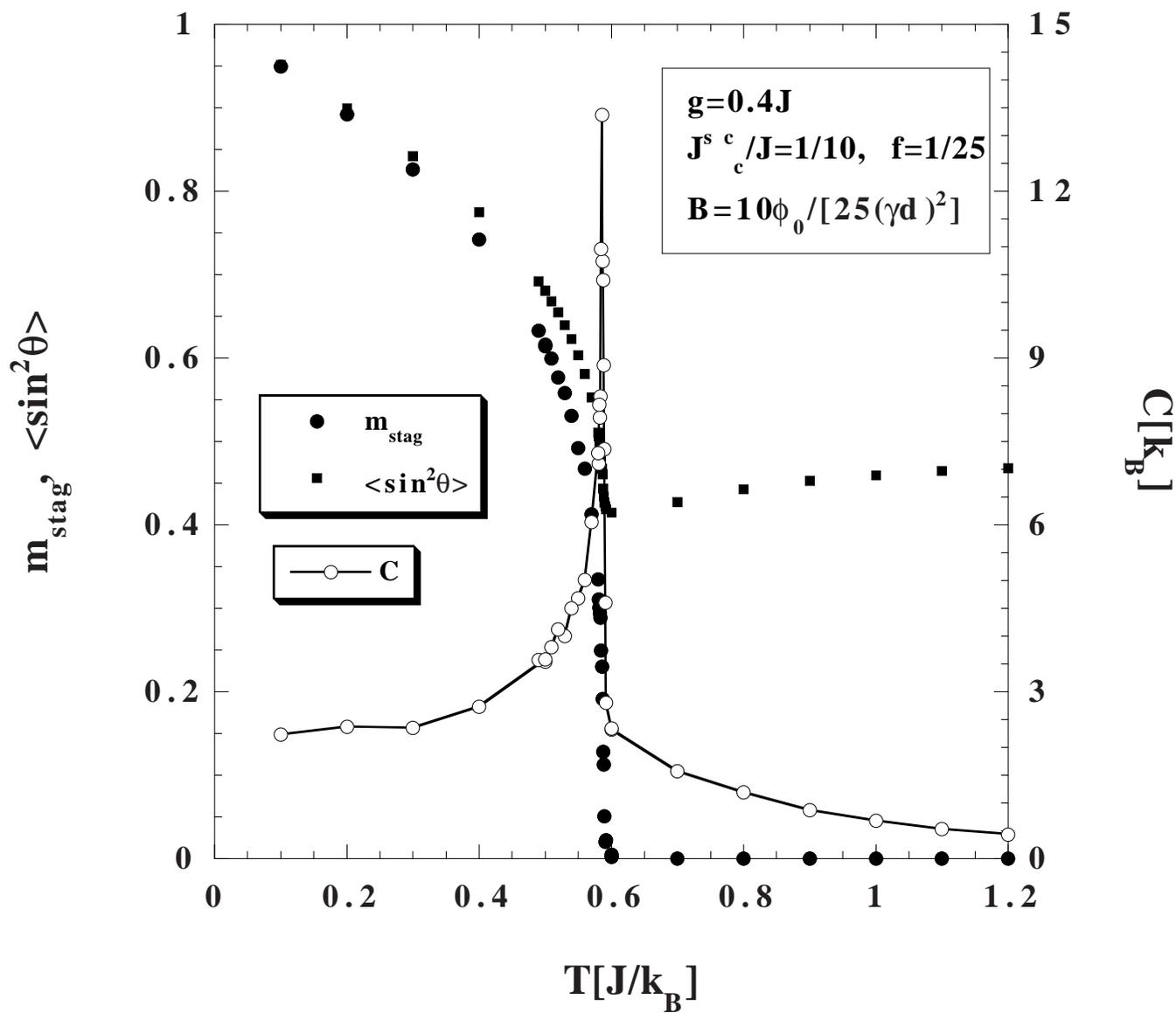
\end{document}